\documentclass[jgrga]{agutex}

  \usepackage{graphicx}
\DeclareGraphicsExtensions{.gif,.eps,.png,.tif,.pdf}
\authorrunninghead{COLLARD ET AL.}

\titlerunninghead{OBSERVING OCEAN SWELL FIELDS}

\authoraddr{Fabrice Collard, Collecte Localisation Satellites,
division Radar, 29280 Plouzan\'{e}, France. (Dr.fab@cls.fr)}

\authoraddr{Fabrice Ardhuin,
Service Hydrographique et Oc\'{e}anographique de la Marine, 29609
Brest, France. (ardhuin@shom.fr)}

\authoraddr{Bertrand Chapron, Ifremer,
Laboratoire d'Oc\'{e}anographie Spatiale, Centre de Brest, 29280
Plouzan\'{e}, France. (bertrand.chapron@ifremer.fr)}

\begin{document}

%
%

\title{Routine monitoring and analysis of ocean swell fields using a spaceborne SAR}
%

%
%


\author{Fabrice Collard}
\affil{Collecte Localisation Satellites, division Radar,
Plouzan\'{e}, France}

\author{Fabrice Ardhuin}
\affil{Service Hydrographique et Oc\'{e}anographique de la Marine,
Brest, France}

\author{Bertrand Chapron}
\affil{Laboratoire d'Oc\'{e}anographie Spatiale, Ifremer, Centre
de Brest, Plouzan\'{e}, France}



%
%
%

%
%


\begin{abstract}
Satellite Synthetic Aperture Radar (SAR) observations can provide
a global view of ocean swell fields when using a specific "wave
mode" sampling. A methodology is presented to routinely derive integral properties of the longer wavelength (swell) portion of the wave spectrum from SAR Level 2 products, and both monitor and predict their evolution across ocean basins. SAR-derived estimates 
of swell height, and energy-weighted peak period and direction, are validated against buoy observations, and the peak directions are used to project the peak periods in one dimension along the corresponding great circle route, both forward and back in time, using the peak period group velocity. The resulting real time dataset of great circle-projected peak periods produces two-dimensional maps that can be used to monitor and predict the spatial extent, and temporal evolution, of individual ocean swell fields as they propagate from their source region to distant coastlines. The methodology is found to be consistent with the dispersive arrival of peak swell periods at a mid-ocean buoy. 
The simple great circle propagation method cannot project the swell heights in space like the peak periods, because energy evolution along a great circle is a function of the source storm characteristics and the unknown swell dissipation rate. A more general geometric optics model is thus proposed for the far field of the storms. This model is applied here to determine the  attenuation over long distances.
For one of the largest recorded storms, observations of 15~s period swells are
consistent with a constant dissipation rate that corresponds to a
3300~km e-folding scale for the energy. In this case, swell dissipation is a significant term in the wave energy balance at global scales.
\end{abstract}

%
%

%

\begin{article}

%
%

\section{Introduction}
Storms over the ocean produce long surface gravity waves that
propagate as swell out of their generation area. In deep water,
the wave phase speed $C$ and period $T$ are proportional. As the
phase speed of the dominant waves  $C_p$ does not exceed 1.2 times
the wind speed~at 10~m height $U_{10}$
\citep{Pierson&Moskowitz1964}, the longest period waves must be
generated by very intense winds. For example, the generation of
waves of period $T$ larger than 16~s requires winds with speeds
over 18 m~s$^{-1}$ blowing over a distance of the order of
1000~km, to produce a significant energy, or yet stronger winds
over a shorter fetch \citep{Munk&al.1963}. Such a large region of
high winds is generally  associated with a smaller storm center
from which the long swells radiate. Waves further evolve after their generation with an important transfer of energy towards both high and low frequencies, due to nonlinear wave-wave interactions. Away from that core, nonlinear interactions become negligible \citep{Hasselmann1963b} and long
period swells have been observed to propagate over very large
distances, up to half-way around the globe \citep{Munk&al.1963},
radiating a large amount of momentum and energy across ocean
basins.  This measurable long-distance propagation is made
possible by a limited loss of energy.

The wave field at any time $t$, latitude $\phi$ and longitude
$\lambda$, is described by its spectral densities $G$ as a
function of frequency $f$ and direction $\theta$. In the limit of
geometrical optics, the spectral density is radiated at the group
speed $C_g$ in the direction of wave propagation, and can be
expressed as a function of $G$ at any previous time $t_0$.
Allowing for a spatial decay at a rate $\mu$,  the spectral energy
balance is
 \cite[e.g.][]{Munk&al.1963} 
\begin{equation}
G(t,\phi,\lambda,f,\theta)= G\left(t_0,\phi_0,\lambda_0
,f,\theta_0 \right) \exp \left(\int_{t_0}^t  -\mu C_g\mathrm{d}t
\right) \label{2Ddispersion}.
\end{equation}
In deep water without current, the initial position $\phi_0$,
$\lambda_0$ and direction $\theta_0$ are given by following the
great circle that goes through the point of coordinates
$\phi,\lambda$ with a direction $-\theta$ over a distance
$X=(t-t_0)C_g=(t-t_0)g/(4 \pi f)$. This corresponds to a spherical
distance $\alpha = X/R$ along the great circle, where $R$ is the
Earth radius.

Equation (\ref{2Ddispersion}) can be used to invert $\mu$ from
wave measurements.   For swell periods shorter than 13~s,
\cite{Snodgrass&al.1966} have measured an e-folding scale
$L_e=1/\mu= 5000$~km  (this number corresponds to a 0.1 dB/degree
attenuation in their analysis). For larger periods,
\cite{Snodgrass&al.1966} could only conclude that $L_e$ is larger,
possibly infinite. In the past 40 years, little progress has been
made on these conclusions \citep{WISE2007}. Yet this question if
of high practical importance, either for wave forecasting
\citep[e.g.][]{Rascle&al.2008} or other geophysical investigations
regarding air-sea fluxes or microseismic noise
\citep[e.g.][]{Grachev&Fairall2001,Kedar&al.2008}.

A theoretical upper bound for $L_e$ is given by the viscous theory
\citep[][ see also Appendix A for a simple derivation]{Dore1978}. According to theory, the largest shears are found
right above the water surface, and the air viscosity dominates the
dissipation of swells, giving, in deep water,
\begin{equation}
L_{e,\max}= \frac{\rho_w g^2}{4 \rho_a \sigma^3 \sqrt{2 \nu_a
\sigma}},
\end{equation}
where $\nu_a$ is the air viscosity, $\sigma = 2 \pi/T$. For
$T=13$~s this gives $L_{e,\max}=45000$~km, which means that over a
realistic propagation distance of 10000~km the energy of 13~s
swells is only reduced by 25\%.

Swells are thus expected to be very consistent over distances that
are only limited by the size of ocean basins. The analysis of
swells at this global scale should provide insights into their
dynamics, including propagation and dissipation, but also into the
structure of the generating areas, in a way similar to the use of
the cosmic microwave background for the analysis of the early
universe. 

The present paper provides two important intermediate
steps toward that goal. First, we demonstrate in section 2 how
sparse data from a single space-borne synthetic aperture radar can
be combined dynamically to provide a consistent picture of swell
fields. This internal consistency reveals the quality of the
SAR-derived dataset which we further verify quantitatively with
buoy data. In section 3, we discuss and derive the asymptotic
far-field swell energy evolution. Numerical investigations are
performed to check the validity of the asymptotic solutions. This
result provides a tool to interpret measured swell heights in
terms of propagation and dissipation. This method is illustrated
with one example that corresponds to a strong swell dissipation.
Conclusions follow in section 4.

\section{Space-time consistency of space-borne swell observations}
Investigations by \cite{Holt&al.1998} and
\cite{Heimbach&Hasselmann2000} have shown that space-borne
synthetic aperture radar (SAR) data can be used to image the same
swell field over 3 to 10 days as it propagates along the ocean
surface. These preliminary studies have shown that the combination
of SAR data at different places and times yields a position of the
generating storm, and predictions for the arrival time of swells
with different periods and directions.
\cite{Heimbach&Hasselmann2000} have further pointed to
shortcomings in the wind provided by an atmospheric circulation model for a given Southern Ocean storm,
based on systematic biases in wind-forced wave model results
compared to SAR observations. Unfortunately, the systematic
analysis of such data has been very limited, and generally
confined to data assimilation in wave forecasting models [e.g.
Hasselmann \textit{et al.} 1997; Breivik \textit{et al.} 1998;
Aouf \textit{et al.}
2006\nocite{Hasselmann&al.1997,Breivik&al.1998,Aouf&al.2006}].
This narrow use of SAR data is due to three essential difficulties.

First, a SAR image is not a picture of the ocean surface and the
relationship between the spectrum of the SAR image and that of the
ocean surface elevation is nonlinear and fairly complex
\citep[e.g.][]{Krogstad1992}. Sophisticated methods have been
developed in order to estimate the surface elevation spectrum
\citep[e.g.][]{Hasselmann&al.1996,Schulz-Stellenfleth&al.2005}.
These methods had to be implemented by the user of the data, and
generally required some \textit{a priori} first guess of the wave
field provided by a numerical model. For longer wave systems, the
imaging mechanisms are essentially quasi-linear, making possible a
simpler methodology used by the European Space Agency (ESA) to
generate a level 2 (L2) product. The method is fully described by
\cite{Chapron&al.2001b}. It uses no outside wave information, and
builds on the use of complex SAR data developed by
\cite{Engen&Johnsen1995} to remove the 180$^\circ$ directional
ambiguity in wave propagation direction. The quality of the L2
data has been repeatedly analyzed
\citep[e.g.][]{Johnsen&al.2006,Collard&al.2005}. Because long
ocean swells have large wavelengths and smaller steepnesses, the
L2 products corresponding to this spectral range have higher
relative quality, confirming that the imaging mechanism is well
described under the quasi-linear assumption.

All SAR data used here are such L2 products, provided by ESA and
obtained with the L2 processor version operational since November
2007, and described by \cite{Johnsen&Collard2004}. The data for
times before that date were reprocessed with this same processor.
In previous real-time data, frequent low wavenumber artefacts were
caused by insufficient filtering of non-wave signatures in the
radar images. This filtering is necessary to remove the
contributions of atmospheric patterns or other surface phenomena
like ships, slicks, sea ice, or islands, with spectral signatures
that can overlap the swell spectra. The L2 product contains
directional wave spectra with a resolution of 10$^\circ$ in
directions and an exponential discretization in wavenumbers
spanning wavelengths of 30 to 800 m with 24 exponentially spaced
wavenumbers, corresponding to wave periods with a 7\% increment
from one to the next. 

The second practical problem is that the data obtained from an
orbiting platform are sparse and with a sampling that makes a
direct analysis difficult. Hereafter, we show that the space-time
consistency of the swell field can be used to fill in the gaps in
the observations and produce continuous observations of swell
periods and directions in space and time.

Third, and last, for a simple use of SAR
data, some parameters that are not affected by the variable resolution is SAR scenes
\citep{Kerbaol&al.1998}. \cite{Schulz-Stellenfleth&al.2007} have proposed to produce parameters representing the entire sea state by extending the resolved spectrum with an empirical windsa contribution. Here we take the opposite approach and restrict the resolved part of the spectrum by using a spectral partitioning (see Appendix B for details) to
retrieve the swell significant wave height $H_{ss}$, defined as four times the square root of the  energy of one swell system, and the peak period $T_p$ and peak
directions $\theta_p$. Thus one SAR typically produced one or two swell parameters for distinct swell systems. 

For $H_{ss}$, only very limited validations have been performed \citep{Collard&al.2006}. We thus perform a thorough analysis of SAR-derived $H_{ss}$ comparing to co-located buoy data (see Appendix B for details). 

The bias on $H_{ss}$ derived from ESA Level 2 products is found to be
primarily a function of the swell height and wind speed,
increasing with height and decreasing with wind speed. Variations
in standard deviation are dominated by the swell height and peak
period, with the most accurate estimations for intermediate
periods  of 14 to 17~s.For wind speeds in the range 3 to 8 m~s$^{-1}$, $H_{ss}$ has a bias of 0.24~m and the standard
deviation of the errors is 0.29~m. 

We thus corrected the  values of $H_{ss}$ by subtracting a
bias model given by
\begin{equation}
b_h=0.11 + 0.1 H_{ss}-0.1 \max\{0,U_{10SAR}-7\}\label{Hserr1}
\end{equation}
where $H_{ss}$ is in meters and the wind speed $U_{10SAR}$ is in
m~s$^{-1}$. 
From now on, all the reported values of $H_{ss}$ will be corrected using this expression. 
After correction, the standard deviation of
$H_{ss}$ estimates is reduced to less than
\begin{equation}
\sigma_h=0.10+\min\left\{0.25 H_{ss},0.8\right\}\label{Hserr2}
\end{equation}
where $\sigma_h$ and $H_{ss}$ are in meters.

The quality of $T_p$ and $\theta_p$ has already been carefully studied and are routinely monitored. The 
root mean square (r.m.s.) error on $T_p$ is less than 10\% of the measured value for $T_p > 12$~s, and the r.m.s. error on $\theta_p$ is 22$^\circ$ for these same swells \citep{Johnsen&Collard2004}, with little bias. Because few directional in situ measurements are available, we demonstrate here an original semi-quantitative dynamical validation of these two parameters. 

\subsection{Virtual wave observers}
Given these SAR-derived estimates of $\theta_p$ and $T_p$,  linear
dispersion relationship and the principle of geometrical optics
can then be exploited to predict arrival times and locations of
the swell.

\begin{figure}[htb]
 \vspace{9pt}
 \centerline{\includegraphics[width=\columnwidth]{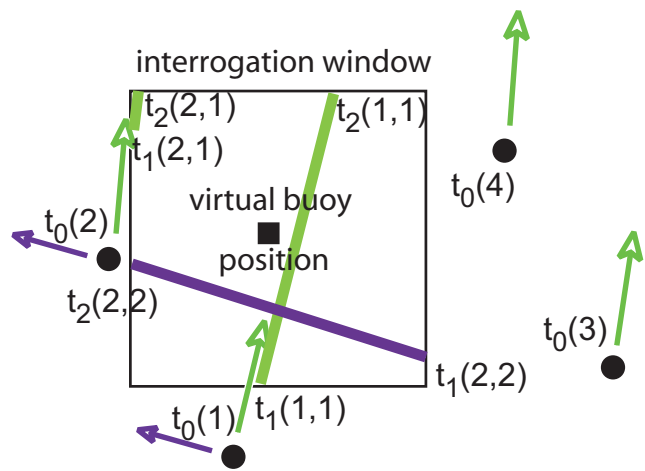}}
 \caption{Schematic definition of a virtual wave observer. Any SAR observation $i$ is available at a time $t_0(i)$ on the black dots.
 All swell partitions $(i,j)$ (here indicated by the arrows) are propagated and may cross the interrogation
 window from time $t_1(i,j)$ to $t_2(i,j)$. \label{fig_virtual}}
\end{figure}

\begin{figure}[htb]
 \vspace{9pt}
\centerline{\includegraphics[width=\columnwidth]{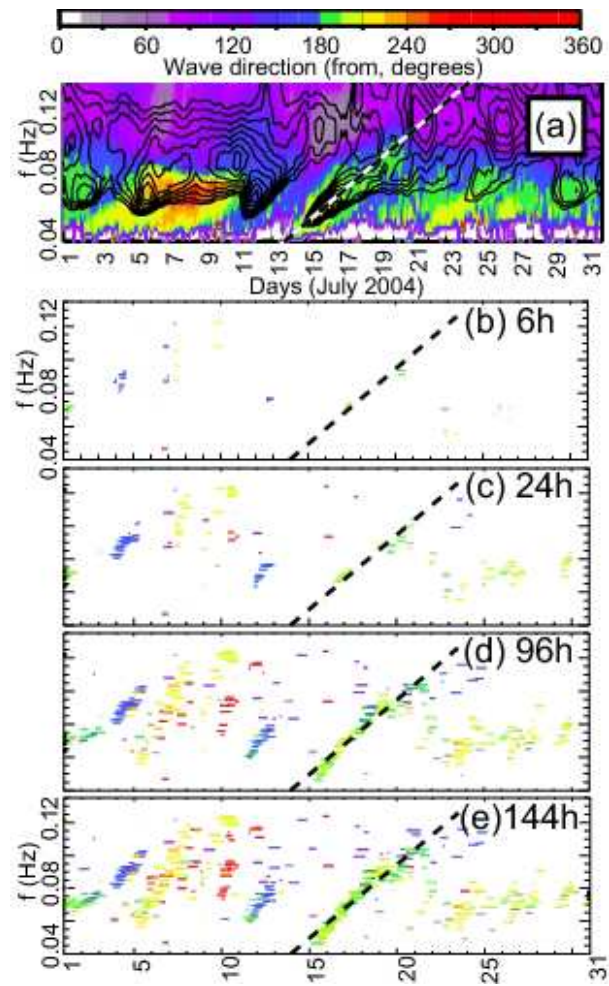}}
 \caption{(a) Energy and mean direction spectrum measured in situ by the Christmas Island buoy (WMO number 51028): contours,
 equally spaced from 0.1 to 1.4, indicate the natural
 logarithm of the spectral energy density $F(f)$. Colors indicate the mean arrival direction at each frequency.
 (b) to (e) Peak direction (colors) as a function of time and peak frequency
 for swell partitions at a SAR virtual buoy located around the Christmas Island buoy (WMO number 51028).
    The maximum propagation time to produce the virtual buoy data is increased from 6 hours (b) to 6 days (e).
    The sloping straight line fitted to the observed SAR ridge from July 16 at 0.05 Hz to July 21 at 0.105 Hz
    is the same as line in (a) that corresponds to the buoy observation. This delayed arrival would correspond
    to a point source at 6100~km from Christmas Island.\label{ridges}}
\end{figure}

In order to obtain swell conditions at the location of a "virtual
wave observer", we define an interrogation window covering 2 by 2 degrees
in latitude and longitude. According to the SAR-derived peak
parameters, swell partitions from the entire ocean basin are
propagated, both forward and backward,  along great circles in
space and time. These theoretical trajectories are followed with a
constant group speed $g T_p/(4 \pi)$, starting off in direction
$\theta_p$ from the observation point. From any observation time
$t_0$, these great circles may cut through the interrogation
window from times $t_1$ to $t_2$ (thick solid lines in figure
\ref{fig_virtual}, with different colors for different
partitions). As the maximum value of $|t_2-t_0|$ and $|t_1-t_0|$
is increased from 6 hours to 6 days, the time-evolution of the
peak frequencies and peak directions at the virtual observer
gradually reveals similar ridges to the one observed in real buoy
measurements (figure \ref{ridges}).

In fig. \ref{ridges}b-d, each horizontal colored segment
corresponds to one swell partition that crosses the spatial window
between times $t_1$ and $t_2$. Some segments are very short,
corresponding to trajectories that barely cut one corner of the
window.

Clearly the SAR detects the direction of the most energetic part
of the wave spectrum measured by the buoys (fig. \ref{ridges}a).
At frequencies above 0.1Hz, the virtual observer patterns appear
rather noisy. Shorter scales are not so correlated.
These shorter components are often observed as part of
the wind sea for which the propagation with a single group speed
and direction is not a good approximation. Also, propagated high
frequency swells, such as the 0.12~Hz waves coming from direction
200 on July 10, do not show up in the real buoy record. This is
possibly the result of a relatively high dissipation rate for
these swells.
For frequencies below 0.08~Hz, the virtual observer shows ridge-like
structures similar to those observed \textit{in situ} due to the
dispersive arrivals of swells from remote storms
\citep[e.g.][]{Munk&al.1963,Gjevik&al.1988}. Even the faintest
events are well detected, such as the 0.06~Hz arrival on 23 July,
even though that 15~s swell of 0.5~m is dwarfed by a another 0.8~m and
12~s swell, and a 2~m and 8~s wind sea. Swell detection with the
virtual observer reaches its limits when the swell height is very low,
such as on 3 July with a 20~s 0.3~m swell well detected by the
buoy (the green-orange ridge at 0.05~Hz). Because we use a single
trajectory emanating from one observed swell partition the
relatively small interrogation window can easily be missed after
10000~km of propagation. 

Our discrete propagation technique suffers from a randomized version of the ``garden sprinkler effect'' that, if not corrected for, can create unrealistic flower-like patterns in the far field of storms in numerical wave models that use a discretized spectrum \citep[e.g.][]{Tolman2002a}. Our choice of a single group speed and direction, because a narrow swell spectrum is not resolved by the SAR, produces a discrete wave field (the dots in figure \ref{fig_fireworks}). With the present processing this is smoothed by the finite size of our interrogation window (figure \ref{fig_virtual}). 
An extension of the present technique
could use neighboring group speeds and directions to take into account the
frequency and directional spread of the sea state, which would allow the use of a smaller window. Just like the estimation of propagated wave heights, discussed below, the estimation of a spectra width that cannot be resolved by the SAR may use some further information on the structure of the generating storm. 

Other errors in the present technique can also be
attributed to the SAR processing. In particular, a maximum
value is defined for the transfer function used to obtain the wave
spectrum from the SAR image \citep{Johnsen&Collard2004}. Although
this is designed to prevent the amplification of measurement
noise, long swells such as this 20~s event have very small slopes, and it is likely
that they are underestimated in the wave spectrum due
to this threshold in the processing.

When propagated for 6 days, without any other information than the
peak frequency and direction at the time of observations, the
waves are remarkably consistent with the latest local observations.
For the southern swells arriving at Christmas Island between July
16 and 21 (figure \ref{ridges}), the difference in arrival times
given by the virtual oberver and real buoy is typically less than 12~h.
This is less than 10\% of the maximum time between the SAR
observation and the virtual observer record. This implies that the
accuracy of the peak period estimate for each SAR partition must
also be less than 10\%, consistent with previous validation studies \citep{Johnsen&Collard2004,Johnsen&al.2006}. The consistency of the arrival directions
along the ridges also suggests that the root mean square (RMS)
error in peak direction estimates must be close to 20$^\circ$,
comparable to the  22$^\circ$ RMS difference between mean wave
directions obtained from SAR wave mode and a numerical wave model
for waves with periods longer than 12~s
\citep{Johnsen&Collard2004}. 

Although it cannot replace the spectral resolving power of a buoy, the performance of the virtual observer is therefore comparable or better to that of human 
observers in terms of peak period and direction \citep{Munk&Traylor1947}. The really missing bit 
is a wave height estimate along the swell propagation path. We will show that this may be obtained 
by  estimating the source storm characteristics and the dissipation rate of swell energy.

\subsection{Storm source identification and "fireworks"}
Along the estimated trajectories, virtual observations can further
be produced in a similar fashion. The animation of these
propagated swells confirms the very well organized nature of storm
swells crossing large ocean basins. 
\begin{figure}[htb]
 \vspace{9pt}
\centerline{\includegraphics[width=\columnwidth]{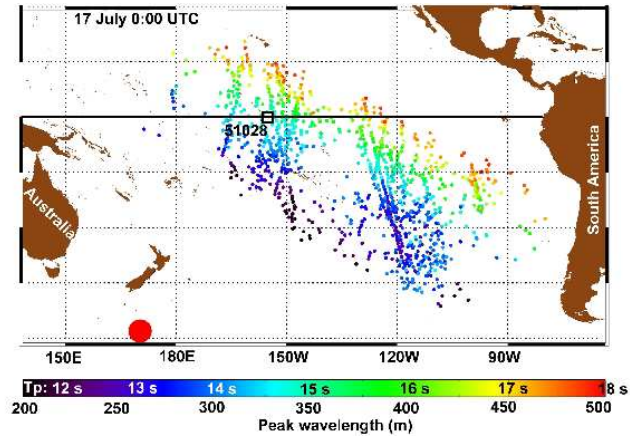}}
 \caption{Snapshot of the "fireworks" animation given in the auxiliary material, for July 17 2004 at 0:00 UTC. Each of the 1071 colored dots represent one observed
 swell partition, within 6 days of its observation, displaced along a great circle with the group speed corresponding to the detected peak period
 in the direction of the detected peak direction. Only swells with tracks that passes within 1000~km of the storm center (red disk) have been
 retained.} \label{fig_fireworks}
\end{figure}
From the relatively sparse and
track-based initial satellite observation sampling, the swell
persistency can then be used to capture "fireworks" patterns
exploding from the few intense storms that occur over a period of
several days (see figure \ref{fig_fireworks} and auxiliary
animation). In large ocean basins where swells are likely to be
imaged several times by the same satellite, these fireworks can be
used to estimate the time of arrival of swells from any given
storm \citep[e.g.][]{Holt&al.1998}. For this reason, these
animations have been produced routinely every day since August
2007, for the Pacific, Indian, and Atlantic oceans (see 
http://www.boost-technologies.com/esa/images/, e.g.
\verb"nrt_pac.gif" for the Pacific Ocean).

\begin{figure}
\noindent\includegraphics[width=0.9\columnwidth]{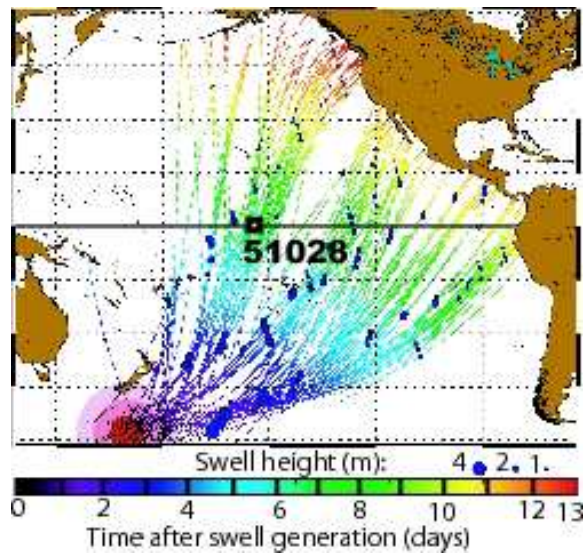}
\caption{Finding the source storm. All swells with a $17\pm 0.5$~s
period that were identified in 13 days of ENVISAT synthetic
aperture radar data over the Pacific, are re-focussed from their
location of observation (filled dots) following their direction of
arrival at the theoretical group speed for 17~s waves. This
focussing reveals a single swell generation event, well defined in
space and time (pink to red disks). The back-tracking trajectories
are color-dated  from black (July 9 2004 18:00 UTC) to red (July
22 2004 18:00 UTC).\label{fig17s}}
\end{figure}

Using backward trajectories, the location and date of swell
sources can further be defined as the spatial and temporal center
of the convergence area and time of the trajectories. These
positions have been verified to correspond to high wind conditions
observed by scatterometers and reproduced by ECMWF wind analyses.
We consider these storms to be the source of all the swell
partitions that produce trajectories that pass within 12 hours and
2000 km of their center. This processing, similar to the one
performed by \cite{Heimbach&Hasselmann2000}, provides a global
view of swell fields in both space and time, extending the
coverage of similar techniques based on buoy data
\citep{Hanson&Phillips2001}. In figure \ref{fig17s}, a swell
covers one Earth quadrant away from the storm, with a large
detection gap that extends from the Southern Pacific to
California. This blank area is the long shadow cast by French
Polynesia where wave energy is dissipated in the surf
\citep[e.g.][]{Munk&al.1963}. Observations were restricted to
swell partitions with periods close to 17~s, but the full dataset
typically covers swells with periods of 12 to 18~s, as shown in
figure \ref{fig_fireworks}.

The apparent self-consistency of both the virtual buoy plot
(figure \ref{ridges}.d) and the fireworks animations, are the
result of the large auto-correlation length of the swell fields, which was
expected from the in situ measurements of
\cite{Darbyshire1958,Munk&al.1963,Snodgrass&al.1966}. Yet, these
plots could not exhibit such patterns without a good
accuracy of the SAR-derived peak periods and directions, used in
the propagation methodology.

\section{Far-field swell energy}
All these illustrations of forward-backward ray tracing indicate
the potential to use a simple Geometrical Optics (GO) strategy.
The next goal is then to determine the strength of the far-field
radiated swell energy. This requires the definition of a swell
source, and an estimation of the swell energy dissipation scale.
For this we define the time $t_0$ as an initial condition after
which there is no significant wave generation or non-linear
evolution, for frequencies less than $f_{\max}$. Namely, at $t_0$
all the wave components with smaller frequencies have already been
generated, so that the radiation of these waves is essentially
fossil and fully governed by geometrical optics. The possible
effects of diffraction and scattering are discussed by
\cite{Munk&al.1963}, and, together with dissipation, will cause
deviations from the G.O. model outlined below. We therefore make no restrictive hypothesis 
on what happens before $t_0$, and thus the motion of the generating storm has no direct effect 
on our results, but it obviously modifies the spatial distribution of the energy at $t_0$, which will be
relevant. 

In reality, swells evolve over the course of their propagation as
the result of their interactions with the local winds, mutual
wave-wave interactions, interactions with other wave systems,
including the local wind sea. Swells are also expected to evolve according to
interactions with other oceanic motions that affect the upper
ocean, namely surface currents, internal waves \citep[e.g.][]{Kudryavtsev1994} and turbulence.
Depth and island scattering effects must also carefully be taken
into account \citep{Snodgrass&al.1966,WISE2007}. Compared to these
different mechanisms, frequency dispersion and angular spreading effects are
certainly the first leading order phenomena to take into account
for the major part of the decrease in the height of the swell
systems. Indeed, as the long swell systems will be characterized
by relatively small steepness parameters, nonlinear mutual
wave-wave interactions do not appear to be important in scattering
surface wave energy more than a few storm radius distances outside
an active generating area \cite{Hasselmann1963b}. Furthermore, the level of turbulence in
the ocean does not appear to significantly affect the waves \cite{Fabrikant&Raevsky1994,Ardhuin&Jenkins2006}, and
the conversion of surface wave energy into internal gravity wave
energy by wave-wave interactions does not seem to be a leading
order sink term for the energy balance of surface gravity waves.
Finally, for the very long trans-ocean fast propagating swell
components, surface current bending effects, proportional to the
ratio between vertical current vorticity and the group velocity,
may also be considered as residual effects. Away from island
obstructions, the ratio between the angular width along the
great-circle observatory points at very large distances from the
generating area, and the mean spread in the generating area is
approximately proportional to $1/\sin \alpha$ with $\alpha$ the
spherical distance from the storm. center The change in spectral
density $F$ defined by
\begin{equation}
F(t,\phi,\lambda, f)=\int_0^{2 \pi} G( t,\phi,\lambda,f,\theta)
d\theta\label{GtoF}
\end{equation}
 follows the spatial expansion (close to the source) and
contraction (as waves approach the antipodes  for $\alpha >
\pi/2$) of the energy front. This transversal dispersion is
associated to a narrowing of the directional spectrum $G$, for
$\alpha < \pi/2$, and a broadening for larger distances.

This approximation applies to large distances, and relatively
small source regions. Closer to the source, the approximation does
not hold. Swell amplitudes radiating from large extended sources
will decrease more slowly than swell amplitudes emanating from
compact sources.

\begin{figure}
\noindent\includegraphics[width=\columnwidth]{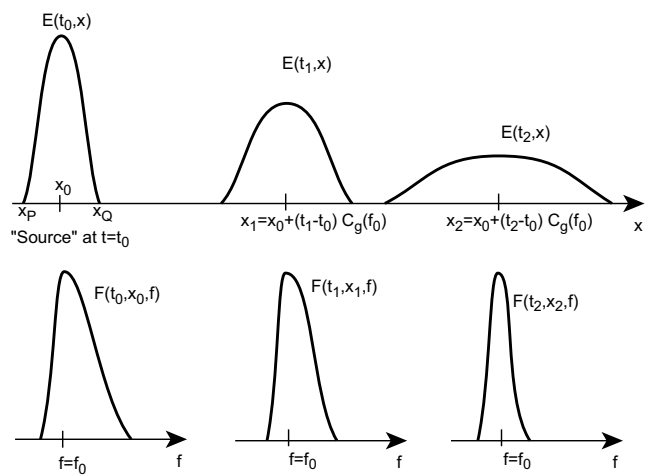}
\caption{Dispersion of linear waves in one dimension. At any given
time the spectrum is given by a propagation of the spectra at
$t=t_0$. Taking $x_1=x_0 + (t_1-t_0) C_g(f_p)$ the spectral
density at $f_p$ is the same as for time $t_0$, but the spectrum
is narrower which gives a smaller elevation variance, $E(t_1,x_1)
< E(t_0,x_0)$.\label{fig1D}}
\end{figure}

Moreover, we can represent swell waves as a linear superposition
of harmonic waves in narrow spectral band. Quite naturally,
through the method of stationary phase, the group velocity is
defined and the slowly-varying wave envelope is found to decay.
This decay is inversely proportional to the square-root of the
distance (figure \ref{fig1D}). Accordingly, far away from the
generating sources, and in the absence of dissipation, the swell
energy
\begin{equation}
E_s(t,\phi,\lambda)=\int_0^\infty F(t,\phi,\lambda,f) {\mathrm
d}f.
\end{equation}
should decrease asymptotically like $1/(\alpha \sin \alpha)$ when
following a wave group (see Appendix C for a detailed proof).

\subsection{The \cite{Snodgrass&al.1966} method}
\begin{figure}
\noindent\includegraphics[width=\columnwidth]{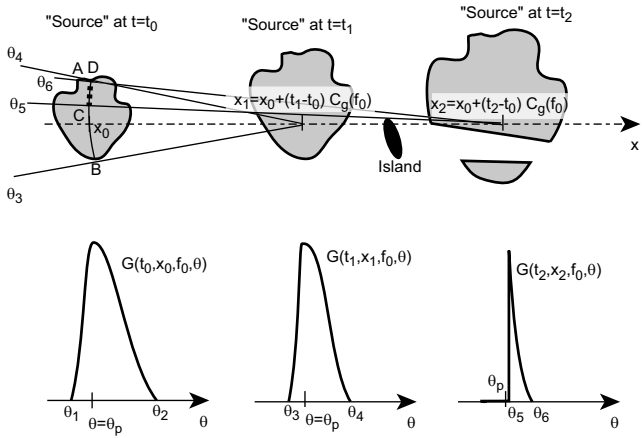}
\caption{Dispersion of linear waves in two dimension, represented
here on a flat surface for simplicity. The variable $y$ has been
dropped as the spectra shown here are all at $y=0$.  We call
"source" the region of the ocean where waves with frequencies
smaller than $f_{\min}$ can be found. As time goes by, the source
expands in space due to both frequency dispersion (like in 1D),
and geometrical dispersion. The wave energy with frequency $f_0$
that will be observed at point $x_1$ (respectively $x_2$) at time
$t_1$ (respectively $t_2$)  is, at time $t_0$, along the thin arc
circle AB (respectively the thick dotted arc circle CD). Due to
the small island between $x_1$ and $x_2$, the energy that would
have been recorded at $x_2$, if the island were not present, is
actually dissipated on the shore of the island. As a result the
local energy density $E(x_2)$ is reduced. At frequency $f_0$,
contributions to $E(x_2)$ only come from angles $\theta_5$ to
$\theta_6$. The directional spectra (bottom) are thus affected by
the blocking effect of islands, and the directional narrowing as
one goes further from $x_0$ (on the Earth this narrowing reverses
after 10000 km of propagation, due to the
sphericity).\label{fig2D}}
\end{figure}

Using measurements with a limited or no directional resolution,
\cite{Snodgrass&al.1966} assumed that wave propagation was
completely blocked by waters shallower than 60 fathoms
(approximately 110~m), and that diffraction could be neglected.
For example, in figure \ref{fig2D}, the island would be
represented by the 60~fathom depth contour. These authors then
estimated a loss of swell energy from the deviation of the ratio
of directionally-integrated spectra, e.g. $F(x_2,f_0)/F(x_1,f_0)$,
compared to what is expected from the lateral dispersion effect,
taking into account islands. Currents, shallow water areas and
diffraction effects around islands are neglected here. These
effects are discussed by \cite{Snodgrass&al.1966}.

Rigorously, their  method is inexact because the recording
stations $x_1$ and $x_2$ measure wave groups that had neither
exactly same propagation directions nor the same position when
they were near $x_0$. Yet, because the wave field in the
neighborhood of $x_0$ is the superposition of many independent
wave trains, one can \textit{assume} that the spectral density $F$
is a smooth function of the direction. Then we may say that over
the intervals
 $\theta_3$ to $\theta_4$ or  $\theta_5$ to $\theta_6$, $G$ does not change so much, i.e. in figure \ref{fig2D},
$G(t_0,\phi_D,\lambda_D,f,\theta_6)\simeq
G(t_0,\phi_D,\lambda_D,f,\theta_5)$. On the sphere the application
of eqs. (\ref{2Ddispersion}) and (\ref{GtoF}) yields
\begin{equation}
F(t,\phi,\lambda,f_0)=\frac{1}{\sin \alpha} \int_C^D G(t_0,
\phi,\lambda,f_0,\theta_0) \frac{{\mathrm
d}s}{R}\left[1+O\left(\frac{1}{\sin \alpha}\right)\right],\label{decayF}
\end{equation}
where the integral is performed over the line segment joining $C$
to $D$. The error relative to the asymptote is the sum of two
terms. One is proportional to the spatial gradient of $F$, due to
the change from the arc circle to the segment, and the other
corresponds to the relative variation of $G$ with $\theta$ over
the range $\theta_5$ to  $\theta_6$, which is small in the far
field, provided that the directional spectrum is smooth enough.

Under these two smoothness assumptions, and for large
$\sin(\alpha)$, the ratios of spectral densities $F$ at $x_1$ and
$x_2$, as used by \cite{Snodgrass&al.1966}, can be compared to the conservative
equations (\ref{decayF}) and used to
diagnose the dissipation of swell energy. This does not apply if
$x_1$ is in the vicinity of the storm or its antipode, where the
observed arrival direction span a large range.

In practice, this method can be very sensitive to the correct
estimation of the island shadowing. For that reason, the
measurement route chosen by \cite{Snodgrass&al.1966} was far from
ideal. Because they needed land to install most recording stations
for the wave measurements, they used an almost north-south great
circle that extends from the south of New-Zealand (Cape Palisader)
to Alaska (Yakutat), a route peppered with islands in its southern
part, and partially blocked by the Hawaiian chain in its northern
part. Also, storms typically refuse to line up with any
measurement array. Their pre-defined great circle, although
designed to follow a typical Southern winter swell propagation
path, always deviated by some extent form the actual track
followed by the most energetic swells they recorded. For the
Indian ocean storms, this difficulty was reduced by the relatively
narrow range of angles that allows propagation from the Indian to
the Pacific ocean.

\subsection{A method using global swell heights}
Now using an instrument with a global coverage, we can carefully
avoid both problems by choosing propagation paths far away from
the smallest island, and by exploiting only observations well
aligned with the storms.  However, due to the limited  spectral
resolution inherent to the SAR wave mode image size and
processing, we cannot use the spectral distribution $G$ or $F$ of
the energy, and can only use the energy $E_s$ integrated over a
swell partition. We thus need for $E_s$ the equivalent of eq. (\ref{decayF}) for $F$.  

\begin{figure}
\noindent\includegraphics[width=\columnwidth]{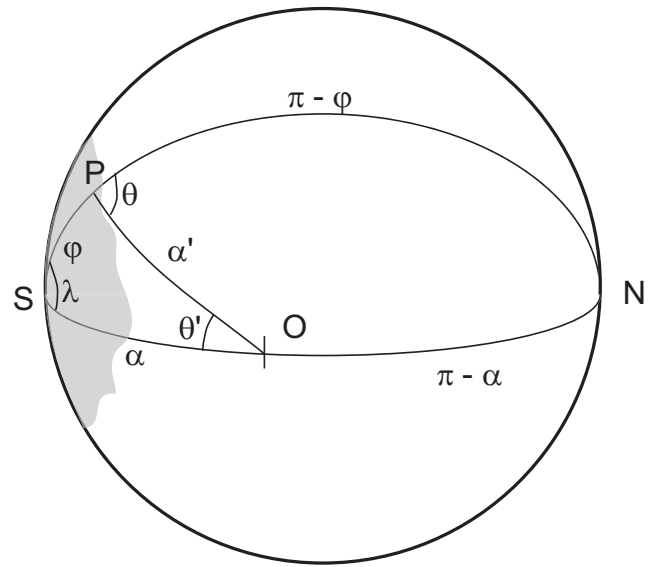}
\caption{Geometry of the ``fossil' swell field distribution at time $t_0=0$ (shaded area) and
observation conditions. Any point $P$ of colatitude $\varphi$ and
longitude $\lambda$ inside of the storm, generates waves that are
observed at point $O$. At time $t$ the observed waves that come
from $P$ have a well defined frequency given by eq.
(\ref{f_sphere}), function of the spherical distance $\alpha'$
between $P$ and $O$, and a well defined direction $\theta$ at $P$,
relative to the North, which gives a direction $\pi-\theta'$ at
$O$. In the triangle $OPS$ the angles $\lambda$, $\theta'$ and
$\pi-\theta$ are related to the distances $\alpha'$,$\varphi$ and
$\alpha$ by the usual spherical trigonometry relationships, e.g.
eq. (\ref{costheta}). \label{fig_sphere}}
\end{figure}

For simplicity, and without loss of generality, we take the source
storm centered at time $t_0=0$ on the pole $S$ defined by a
colatitude $\varphi=\phi-\pi/2=0$, so that the distance from the
storm center is $r=R \varphi$. We consider the swell energy $E_s$
observed at a position defined by the spherical distance $\alpha$
and we take the reference meridian to be in the direction of the
observation point (figure \ref{fig_sphere}).
We will later assume that the source area is relatively small with
a size $R \Delta_\alpha$, where $\Delta_\alpha$ is the maximum
value taken by $\varphi$ (figure \ref{fig_sphere}). In all the
following derivations, we have chosen a fixed frequency $f_0$ and
we follow a wave group of that frequency. The time of observation
$t$ is thus related to $\alpha$ by $t=R\alpha/Cg(f_0)$, so that
the variable $t$ will be omitted.

In appendix C, we prove and verify that, in the absence of dissipation  the swell 
energy  $E_s$ decreases like  $1/(\alpha \sin \alpha)$ for large values of $\alpha$, and that typical storms should 
produce swells within 20\% of this asymptote at distances larger than 4000~km from the storm center. 
A much larger observed deviation should thus reflect a gain or loss of energy by the propagating swells. 
The expected departures from the asymptotic evolution should be
compared to those due to swell dissipation or generation.  Even
with perfect SAR observations, this is the intrinsic limit of the
present method. A 20\% error in energy conserving conditions may
be misinterpreted as a dissipation or generation with an e-folding
scale of the order of $20000$~km, which gives a
 20\% energy change as
waves propagate from 4000 to 8000~km away from the storm source.

\subsection{Illustration}
To illustrate the method described above, we analyse of one of the
most powerful swell field recorded over the past 4 years by ENVISAT's
ASAR. The swell case illustrated in figure \ref{fig17s} is not
well suited due to the islands in the south-north swell tracks and
the poor sampling of ENVISAT for the tracks going north-east from
the storm, we have thus chosen another source, found on February 12 2007 at
18:00 UTC, and located  at 168 E and 38 N. This swell was generated by a storm moving Eastward
with maximum Westerly winds of 26~m~s$^{-1}$ at the indicated date, and subsiding to less that 22~m~s$^{-1}$ and veering to South-Westerly 12~h later (according to ECMWF 0.5$^\circ$ resolution analyses). The storm motion in the direction of wave propagation certainly helped to amplify the local windsea, with a maximum significant wave height $H_s$ of 14.1~m at 00:00 UTC on February 13 (according to a numerical wave model configuration that is otherwise verified to produce root mean square errors less than 9\% for $H_s > 8$~m).

Using SAR-measured wave periods and directions at different times
and locations, we follow great circle trajectories backwards at
the theoretical group velocity. The location and date of the swell
source is defined as the spatial and temporal center of the
convergence area and time of the trajectories.
\begin{figure}
\noindent\includegraphics[width=\columnwidth]{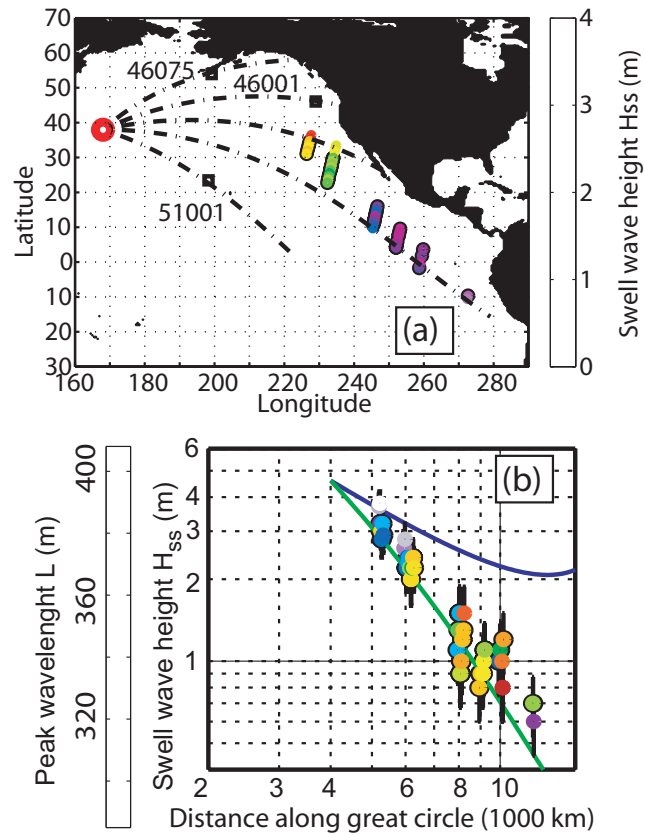}
\caption{(a) Location of SAR observations with a 15~s peak period
swell system corresponding to the 12 February source, with
outgoing directions of 74 to 90$^\circ$. The same swell was also
observed at all buoys from 46075 off Western Alaska, to 51001 in
Hawaii. The dash-dotted line represent great circles leaving the
storm source with directions 42, 59, 74, 90 and 106$^\circ$. (b)
Observed swell wave height as a function of distance. The solid
lines represent theoretical decays using no dissipation (blue), or
a the fitted linear dissipation (green), for swells observed in
February 2007. Circled dots are the observations used in the
fitting procedure. Error bars show one standard deviation of the
expected error on each SAR measurement.}\label{fig_mu}
\end{figure}
We chose a central peak period, here 15~s, and track the swells
forward in space and time, starting from the source center at an
angle $\theta_0$, following ideal geodesic paths in search of SAR
observations. Along each track, SAR data are selected if they are
acquired within 3~hours and 100~km from the theoretical time and
position. Great circle tracks are traced from the source in all
directions, except for angular sectors with islands.

In order to obtain enough SAR data, we repeat this operation for
regularly spaced values of $\theta_0$ with a step of $2^\circ$. In
our case, when varying $\theta_0$ from 74 to 90$^\circ$ (counted
clockwise from North), this procedure produced 58 SAR measurements
with one swell partition that had a peak wavelength and direction
within 50~m and 20$^\circ$ of the expected theoretical value.

If no energy is lost by the wave field, $E_s$ decreases
asymptotically as $1/[\alpha \sin(\alpha)]$ away from the source.
Among the 58 swell observations, we further removed all the data
within 4000 km of the source center, to make sure that the
remaining data are in the far field of the storm, and data with a
significant swell height $H_{ss}$ less than 0.5~m, after bias
correction based on the error model. This makes sure that the
signal to noise ratio in the image is large enough so that the
wave height estimation is accurate enough.

We then have 35 observations for which we assume that $E_s$ is
only a function of $\alpha$, and we define the dissipation rate
\begin{equation}
\mu=-\frac{1}{E_s R}\frac{\mathrm{d} \left(E_s \alpha \sin \alpha
\right)}{ \mathrm{d} \alpha}.\label{alphadef}
\end{equation}
Positive values of $\mu$ correspond to losses of wave energy
(figure \ref{fig_mu}). We then fit an analytical function
$H_{ss}(\alpha)$ to the data, defined by a constant $\mu$ and
$H_{ss}(\alpha=\pi/5)$, i.e. the swell height at 4000~km from the
source. Here the couple $H_{ss}(\alpha=\pi/5)=4.4$~m and
$\mu=3.7\times 10^{-7}$~m$^{-1}$ gives the least square difference
between the decay with constant $\mu$ and the observed swell
decays. Further, the uncertainty of that dissipation rate may be
estimated from the known uncertainty of the SAR measurement of
$H_s$, given by eq. (\ref{Hserr1})-(\ref{Hserr2}). A more simple
error model, with larger errors based on the $H_{s12}$ analysis by
\cite{Johnsen&al.2006}, does not significantly alter this
analysis. Using that error model and neglecting other sources of
error in the present analysis, we perturbed the observed swell
heights independently to produce a 400 ensemble of synthetic data
sets. Taking the 16\% and 84\% levels in the estimation of $\mu$,
that would correspond to one standard deviation for a Gaussian
distribution, we find that $3.1\times 10^{-7}<\mu < 4.0\times
10^{-7}$ at the 68\% confidence level. This is the first ever
estimation of the uncertainty on an observed swell dissipation
rate. These values of $\mu$ are more than twice larger than
reported by \cite{Snodgrass&al.1966} for smaller amplitude swells.

The formidable height of 4.4~m at a distance of  4000~km was
observed by the SAR for all outgoing directions from at least $74$
to $106^{\circ}$. This same swell was also recorded by buoys in
the North-East of Hawaii (NDBC buoy 51001), also with a peak
period of 15~s, and a height of 3.4~m on 16 February 2007 at 0:00
UTC. That buoy is located 3300~km away from the center and in a
direction close to 112$^\circ$. Looking in the North-East quadrant
of the storm, one also finds a trace of the swell at buoy 46005,
off the Washington coast (4900~km in direction 59$^\circ$). There
the swell was also observed with a 15~s period and a maximum
height of 3.2~m on February 17 at 17:00 UTC, similar to the SAR
observations for the same distance. For directions closer to
Northbound, either the generation was weaker or the Alaskan
islands sheltered the coastal buoys. For example, the same swell
was also recorded by NDBC buoy number 46075, off Shumagin Island,
Alaska, at a distance of 3000~km from the source, in the direction
42$^\circ$. At that buoy, the peak period was 15.0~s with a
maximum swell height of 1.3~m on 15 February 2008 at 18:30 UTC.

Thus the power radiated by the storm  is of the order of is 0.5~TW
at 4000~km from the storm center, spread over a 50$^\circ$ angular
sector. This power is about 16\% of the estimated 3.2 TW annual
mean flux that reaches the world's coastlines
\citep{Rascle&al.2008}. However, the observed dissipation rate
corresponds to an e-folding scale of 3300~km for the energy.
Taking an average propagation distance of 8000~km, only 160~GW
would make it to the shore. If the same dissipation rate prevailed
closer to the source, then the power radiated at 1000~km form the
storm center was 1.4~TW.

We thus expect that the far field dissipation of swells, in spite
of the small steepness of these swells, plays a significant role
in the air-sea energy balance. This effect probably explains the
systematic positive bias for predicted wave heights in wave models
that neither account for swell dissipation nor assimilate wave
measurements \citep[see e.g.][]{Rascle&al.2008}.

\section{Conclusions}
Taking advantages of the satellite observations of unprecedented
coverage and quality, investigations can repeat and complement the
pioneering analysis of swell evolution performed in the 1960s.
Severe storms can generate relatively broad spectra of large
surface waves. But rapidly, the redistribution of energy, through
linear dispersion and nonlinear interaction mechanisms, becomes
very effective. The initial wind waves become swells outrunning
the wind, leading to the apparition long-crested systems. The
propagation properties of these surface gravity waves have been
found to closely follow principles of geometrical optics. The
consistent patterns of swell fronts dispersing over thousands of
kilometers was shown to be useful to provide time series at
"virtual wave observing stations", filling gaps in space and time in between the
orbit cycles of observation. When compared to buoy measurements, the
present results give an explicit dynamical validation of
the SAR-derived spectral parameters. As the speed of waves in deep
water is proportional to their period or wavelength, information
carried by the SAR-derived period and direction distributions,
observed at a fairly large distance from the generating area,
pertains to the wind conditions existent up to 15 days before.

We also discussed how the swell energy should, in the absence of
dissipation, decay in the far field of the storm like $1/(\alpha
\sin \alpha)$ where $\alpha$ is the spherical distance between the
storm center and observation point. Exploiting that property
allowed us to estimate a dissipation rate $\mu$ of swell energy
with unprecedented accuracy, establishing that swell dissipation
can be a significant term in the global wave energy budget.

The proposed methodolgy performed here requires data far enough
from the source, typically more than 4000~km, in order to approach
this simple asymptotic behavior. At the same time, the swell
amplitude should be large enough to be accurately measured by the
SAR. Some knowledge of the spectral shape and its spatial
distribution inside the storm may be useful to provide better
estimates of $\mu$ for low dissipation cases, or closer to the
storm centers. These further analyses will likely benefit from the
joint use of data from altimeters, SARs, and other sources of
spectral wave information.

A more systematic analysis and interpretation of this dissipation
will be reported elsewhere \citep{Ardhuin&al.2009b}, with
applications to wave forecasting models
\citep{Ardhuin&al.2008d,Ardhuin&al.2009}. The parameterization of
the dissipation rate could also be used to produce a data-based
forecasting system, extending our virtual buoy technique to the
estimation of swell heights, with a forward propagation of
observations.

Going in the opposite direction, toward the storm source, it is
possible that the analysis of swell fields could provide a "new"
way of looking into the poorly observed structure of severe
storms. Because the usual remote sensing techniques for estimating
wind fields either do not work for very high winds or are not well
validated \citep[e.g.][]{Quilfen&al.2006,Quilfen&al.2007}, the use
of far-field swell information may provide an interesting
complement to the local wind speeds and wave heights. The is inverse problem 
has already been formulated by  \cite{Munk&al.1963} who already proposed an elegant
heterodyning technique to push the spatial resolution for the
estimation of storm location from swell data, while
\cite{Heimbach&Hasselmann2000} have proposed to use wave models to
correct wind field errors. The quality of the SAR-derived swell
parameters that are coming out of today's ENVISAT and tomorrow's
Sentinel-1, together with a good understanding of the swell energy
budget, including its dissipation revealed here, may finally
enable this vision.


%
%
\appendix \section{Viscous theory for air-sea interaction}
\begin{figure}[htb]
 \vspace{9pt}
\centerline{\includegraphics[width=\columnwidth]{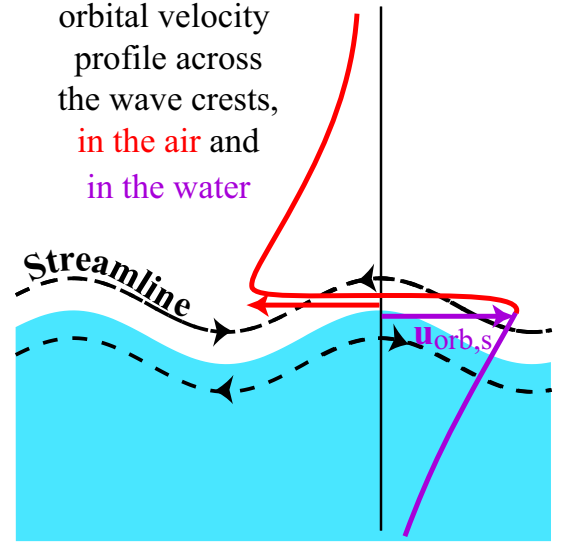}}
 \caption{Boundary layer over waves in the absence of wind. Because of the larger inertia
 of the water compared to the air, most of the adjustment from the sub-surface velocity to the
 free stream velocity in the air occurs on the air-side of the surface.} \label{figBL}
\end{figure}

 For the sake of simplicity we will consider here the case of
monochromatic waves propagating in the $x$ direction only, and we
will neglect the curvature of the surface. For the small steepness
swells considered here that latter approximation is well founded
and a more complete analysis is given by Kudryavtsev et Makin (2004)\nocite{Kudryavtsev&Makin2004}. For deep water waves, the free stream velocity above the waves, just outside of
the boundary layer is $u_{+}(x,t)=- \sigma a \cos(k x -\sigma t)$,
where $a$ is the swell amplitude and $\sigma = 2 \pi/T$ is the
radian frequency. The sub-surface velocity is $u_{-}(x,t)=\sigma a
\cos(k x -\sigma t)$  (figure \ref{figBL}). Due to the
oscillations that propagate at the phase velocity $C$, the
horizontal advection of any quantity $X$ by the flow velocity $u$,
given by $u
\partial X /\partial x$, can be neglected compared to its rate of change in time
$\partial X /\partial t$ since the former is a factor $u/C$ smaller than the
latter, which is typically less that 0.1 for the swells considered here.
Defining $\widetilde{u}(x,z,t)=\left<u(x,z,t)\right>-u_{-}(x,t)$, where the
brackets denote an average over flow realizations for a given wave phase. The horizontal momentum
equation is thus approximated by,
\begin{equation}
\frac{\partial \widetilde{u}}{\partial t}=-\frac{1}{\rho_a}\frac{\partial
p}{\partial x} - \frac{\partial u_-}{\partial t} + G
\end{equation}
where $G$ represents the divergence of the vertical viscous and turbulent
fluxes of horizontal momentum,
\begin{equation}
G=\nu \frac{\partial^2 \widetilde{u}}{\partial z^2} + \frac{\partial
\left<u'w'\right>}{\partial z}.
\end{equation}

Because the boundary layer thickness $\delta$ is small compared to the
wavelength, the pressure gradient in the boundary layer is given by the
pressure gradient above the boundary layer, in balance with the horizontal
acceleration. This is another way to write Bernoulli's equation (e.g. Mei 1989\nocite{Mei1989}),
\begin{equation}
-{\partial p}/{\partial x}/{\rho_a}=-\sigma^2 a \sin(k x -\sigma
t)={\partial u_+}/{\partial t}.
\end{equation}

 This yields
\begin{equation}
\frac{\partial \widetilde{u}}{\partial t}=2 \frac{\partial u_+}{\partial t} + G
\end{equation}
with the boundary condition for $z \gg \delta$, $ \widetilde{u}$ goes to $2
u_+(x,t)$. The equation for the horizontal momentum is thus exactly identical
to the one for the oscillatory boundary layer over a fixed bottom with wave of
the same period but with an amplitude twice as large. In the viscous case, $G=\nu {\partial^2 \widetilde{u}}/{\partial z^2}$ and one
recovers, after some straightforward algebra, the known viscous result, i.e.,
for $z
> \zeta$,
\begin{equation}
\widetilde{u}(x,z,t)=2 \sigma a \left[\mathrm{e}^{-z_+} \cos\left(k x - \sigma t
-z_+\right) - \cos\left(k x - \sigma t \right)\right] + O
(\rho_a/\rho_w)\label{uvisc}
\end{equation}
where $z_+=(z-\zeta)/\sqrt{2 \nu/\sigma}$, with the surface elevation
$\zeta(x,t)=a \cos(k x -\sigma t)$. 

The dissipation rate of energy is given by the mean work of the 
viscous stresses,
\begin{equation}
\beta = C_g \mu_v = \frac{\left<\rho_a \nu u \partial u/\partial z\right>}{\rho_w g a^2/2}.
\end{equation}
Equation (\ref{uvisc}) gives, 
\begin{equation}
\frac{\partial \widetilde{u}}{\partial z}=- 2 \sigma a  \frac{\mathrm{e}^{-z_+} }{\sqrt{2 \nu / \sigma}} \left[\cos\left(k x - \sigma t
-z_+\right) - \sin\left(k x - \sigma t
-z_+\right)\right] 
\end{equation}
This gives the
low frequency asymptote to the viscous decay coefficient \citep{Dore1978}, 
\begin{equation}
\mu_v=-2 \frac{\sigma^2}{g C_g} \frac{\rho_a}{\rho_w}\sqrt{2
\nu \sigma}\rho_a/\rho_w .
\end{equation}
This result was previously obtained using a
Lagrangian approach without all the above simplifying
assumptions \citep[][their equation 7.1 in which our $\mu$ correspond to their 
$2 \alpha$]{Weber&Forland1990}. The full viscous result is obtained by
also considering the water viscosity $\nu_w$, which gives the $O
(\rho_a/\rho_w)$ correction for the motion in the air, and the classical
dissipation term with a decay $\mu_{vw}=-4 k^2 \nu_w /C_g$, which dominates
for the short gravity waves. 

The total disspation rate is simply $\mu=\mu_v + \mu_{vw}$. 
For a clean surface with $\nu_w=3\times 10^{-6}$m$^2$~s$^{-1}$, 
$\nu=1.4\times 10^{-5}$~m$^2$~s$^{-1}$, and $\rho_a/\rho_w=0.0013$, the two terms 
are equal for waves with a period $T=1.3$~s and a wavelength of 2.6~m. The air viscosity 
dominates for all waves longer than this, which is typically the range covered by spectral 
wave models for sea state forecasting. 

For a comparison with fixed bottom boundary layers,
the Reynolds number based on the orbital motion should be
redefined with a doubled velocity and a doubled displacement, i.e.
Re$=4 u_{\mathrm{orb}} a_{\mathrm{orb}}/\nu$. For monochromatic
waves $a_{\mathrm{orb}}=a$ and $u_{\mathrm{orb}}=a \sigma=2 \pi
a/T$. For random waves, investigations of the ocean bottom
boundary layer suggest that the boundary layer properties are
roughly equivalent to that of a monochromatic boundary layer
defined by significant properties (Traykovski et al. 1999\nocite{Traykovski&al.1999}).

Although the wind was neglected here, it should influence the
shear stresses when its vertical shear is of the order of the
wave-induced shear. Taking a boundary layer thickness $\delta$ and
wind friction velocity $u_\star$, and assuming a logarithmic wind
profile, this should occur when $u_\star/(\kappa \delta)$ exceeds
$2 u_{\mathrm{orb}} / \delta$, where $\kappa$ is von
K{\'a}rm{\'a}n's constant. This corresponds to, roughly, $u_\star
> u_{\mathrm{orb}}$. For swells with T $< 15$~s and $H_{ss} > 2$~m
(i.e. $u_{\mathrm{orb,s}} > 0.4$~m~s$^{-1}$), and winds less than
7~m~s$^{-1}$ (i.e. $u_\star < 0.2$~m~s$^{-1}$), the wind effect on
$f_e$ may be small and the previous analysis is likely valid.  In
general, however, the nonlinear interaction of the wave motion and
wind should be considered, which requires an extension of existing
theories for the distortion of the airflow to finite swell
amplitudes.

Finally, the above result for the air viscosity is easily generated to any water depth $D$ by dividing the 
free stream velocity $u_{+}(x,t)$ by $\tanh(kD)$, so that the dissipation rate is a factor 
$1/\tanh^2(kD)$ larger in intermediate water depth. 

\section{Quantitative validation of $H_{ss}$}
A classical analysis of SAR estimation errors is provided by a
direct comparison of swell parameters, estimated from level 2
products, with buoy measurements at nearly the same place and time
\citep{Holt&al.1998, Johnsen&Collard2004}.

Previous validations were presented for the total wave height
$H_s$ \citep{Collard&al.2005} or a truncated wave height $H_{s12}$
defined by chopping the spectrum at a fixed frequency cut-off of
$1/12$~Hz. For that parameter, \cite{Johnsen&Collard2004}
 found a root mean square (RMS) difference of
0.5~m, when comparing SAR against buoy data, including a bias of
0.2~m. In the present study, we use $H_{ss}$ values obtained from
both SAR and buoy spectra. 

For each wave spectrum observed in the world ocean, swell
partitions are extracted providing estimations of $H_{ss}$, $T_p$,
and $\theta_p$. In practice, the L2 spectra are first smoothed
over 3 direction bins (30$^\circ$ sectors) and 3 wavenumber bins, in order to remove multiple peaks that actually correspond to the same swell system.
The peaks are then detected and the energy associated to each peak
is obtained by the usual inverted water-catchment procedure
\citep{Gerling1992}. The swell peak period is defined as the
energy-weighted average around $\pm 22$\% of the frequency with
the maximum energy. Likewise the peak direction
$\theta_p$ is defined as
      the energy-weighter direction within 30$^\circ$ of the peak
      direction.

A preliminary validation of $H_{ss}$
was performed by \cite{Collard&al.2006}, using L2 processing
applied to 4 by 4 km tiles from narrow swath images exactly
located at buoy positions. That study found a 0.37~m r.m.s. error.
This smaller error was obtained in spite of a 4 times smaller
image area that should on the contrary produce larger errors due to statistical uncertainties. This suggests that a significant part of the "errors" in
SAR validation studies are due to the distance between SAR and
buoy observations.

The swell height validation has been repeated here, using all buoy
data from 2004 to 2008, located within 200 km and 1 hour of the
SAR observation. These co-located data are made publically
available as part of the XCOL project on the CERSAT ftp server,
managed by Ifremer. Because we wished to avoid differences due to
coastal sheltering and shallow water effects, we restricted our
choice of buoys to distances from the coast and the 100~m depth
contour larger than 100~km. As a result, most selected buoys are
not directional, and we use partitions in frequency only. For the
present validation, differently from other section of this paper,
we thus define a swell partition as the region between two minima
of the frequency spectrum. The corresponding energy $E_s$ gives
the swell height $H_{ss}=4\sqrt{E_s}$. The buoy swell height is
then defined from the energy contained within the frequency band
of the SAR partition. The peak period is then estimated as the
period where the buoy spectrum is maximum. The database includes
15628 swell partitions observed by the SAR, with matched buoy
swell partitions.

Many of these observations correspond to relatively short swells,
for which the waves are poorly imaged. We have thus defined a
subset of the database by imposing the following conditions. First
the image normalized variance, linked to the contrast intensity
and homogeneity, should be in the range 1.05 to 1.5, which limits
the dataset to 6651 observations. This removes SAR data with
non-wave features (slicks, ships ...) that would otherwise
contaminate the wave spectra. Second, both the SAR and buoy peak
periods are restricted to the 12 to 18~s range, which reduces the
dataset to 4136 observations, and removes most of the problems
related to the azimuthal cut-off. Third and last, the SAR-derived
wind speed $U_{10SAR}$ is limited to range from 3 to 9 m~s$^{-1}$
in order to remove low winds with poorly contrasted SAR images and
high winds which may still cause some important azimuthal cut-off
and contamination of swell spectra by wind sea spectra. This gives
subset A, with 2399 observations. The resulting heights are compared to buoy
measurements in figure 10.
\begin{figure}
\noindent\includegraphics[width=\columnwidth]{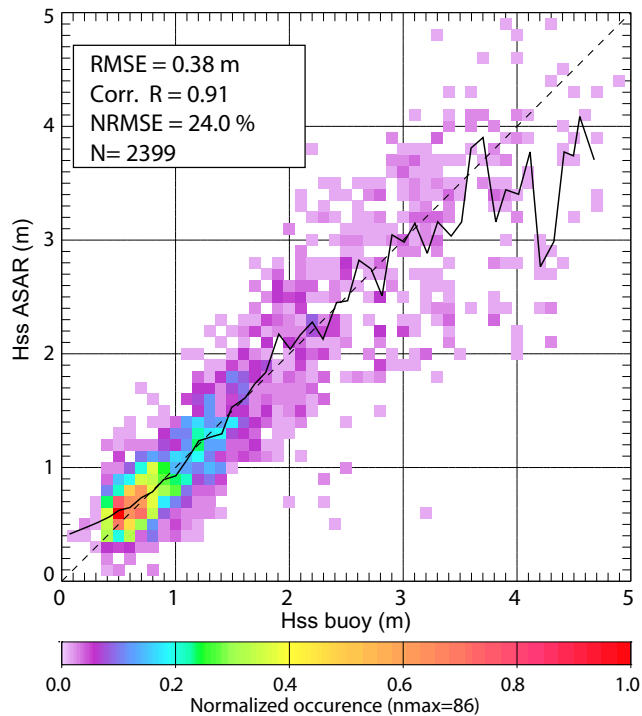}
\caption{ASAR-derived swell partition heights versus buoy swell
partition heights after bias correction using eq. (\ref{Hserr1}),
for subset A. The solid line joins the median values from SAR
observations in each 0.1~m class of buoy-measured height.
\label{figscat}}
\end{figure}

When the maximum wind is reduced to 8~m~s$^{-1}$, giving subset B,
the differences between SAR and buoy data is reduced, with further
reductions when the maximum distance between SAR and buoy data is
reduced from 200 to 100~km to give subset C (table 1).

%
\begin{table}[tb]
\caption{Error statistics for swell partitions heights and peak
periods derived from SAR wave mode data (after bias correction)
against buoy-derived data, for subsets A, B and C of the
co-located database. Subset A contains 2399 observations. Subset B
is restricted to $U_{10SAR} \le  8$m/s and contains 1936
observations. Subset C is further restricted to SAR-buoy distances
less than 100 km, and contains only 460 observations. RMSE stands
for root mean square difference, while the NRMSE is the RMSE
normalized by the root mean square observed value. The scatter
index (S.I.) is equivalent to the NRMSE computed after bias
removal. Finally, $r$ is Pearson's linear correlation
coefficient.} \vspace{5pt}
\begin{tabular}{lcc}
 & $H_{ss}$ & $T_{p}$  \\
\tableline &  &     \\
subset A, bias   &   0.00 m    &      0.27 s   \\
subset A, RMSE   &   0.38 m    &     1.14   \\
subset A, S.I.   &   24.0\%    &     7.9\%  \\
subset A, NRMSE  &   24.0\%    &     8.2\%  \\
subset A, $r$      &   0.91      &    0.61    \\
\tableline &   &      \\
subset B, bias   &   0.00 m     &      0.24 s  \\
subset B, RMSE   &   0.35 m     &      1.11 s  \\
subset B, S.I.   &   23.5\%    &      7.8\%  \\
subset B, NRMSE  &   23.5\%    &      8.0\%  \\
subset B,  $r$     &   0.92      &     0.62    \\
\tableline &  &   \\
subset C, bias   &   0.02 m    &      0.32   s  \\
subset C, RMSE   &   0.29 m     &      1.07  s   \\
subset C, S.I.   &   22.4\%    &      7.3\%    \\
subset C, NRMSE  &   22.5\%    &     7.7\%     \\
subset C, $r$      &   0.92      &     0.64      \\
\tableline &   &   \\
\end{tabular}
\end{table}

\section{Derivation and verification of the asymptotic swell energy without dissipation}
\subsection{Derivation}
The swell energy $E_s$ is an integral of the local spectrum $G$ over both
frequencies $f$ and arrival directions $\theta$,
\begin{equation}
E_s(\alpha)= \int_0^{2 \pi} \int_0^\infty
G\left(t,\phi',\lambda',f,\theta' \right)\mathrm{d f}\mathrm{d
\theta' }.  \label{ES2}
\end{equation}
Using eq. (\ref{2Ddispersion}) this local integral, can also be
written as an integral over the entire source area $\Omega$. The
spherical distance between any point $P(\varphi,\lambda)$ in the
source region and the observation point $O(\alpha,0)$ is
$\alpha'$. The observed frequency that is due to this source point
is
\begin{equation}
f=gt/( 4 \pi R \alpha')=f_0
\frac{\alpha}{\alpha'}\label{f_sphere}.
\end{equation}
We may replace $f$  by $\alpha'$ in eq. (\ref{ES2}),
\begin{equation}
E_s(\alpha) = \frac{f_0}{\alpha }\int_0^{2 \pi} \int \frac{
\alpha^2 G\left(t_0,\phi,\lambda,f,\theta \right)}{\alpha'^2}
\mathrm{d \theta'}\mathrm{d \alpha'} \label{ES2b}.
\end{equation}
For a circular uniform storm of radius $r$ with isotropic spectra,
as used in figure \ref{fig_asymptote}, $E_s(\alpha)$ is given by
the integral over $\alpha'$ weighted by the directional width of
the spectrum $\Delta \theta'$. That width is given by the
spherical trigonometry relationship
\begin{equation}
\Delta \theta' =2 \theta'_{\max} =2 \arccos \left[\frac{\cos
\varphi_{\max} - \cos \alpha \cos \alpha'}{\sin \alpha \sin
\alpha'}\right], \label{costheta}
\end{equation}
with $\varphi_{\max}=r/R$.

For general spectral distribution, we may transform the
integration variables  $(\alpha',\theta')$ which are the
colatitude and longitude coordinates on the sphere with a pole at
the observation point, to
 coordinates $(\varphi,\lambda)$ with a pole
in the center of the swell field at $t_0$. The transformation Jacobian is simply given
by the equality of the elementary area on the unit radius sphere
$dA= \sin \varphi {\mathrm d} \phi {\mathrm d} \lambda = \left|
\cos \phi\right| {\mathrm d} \phi {\mathrm d} \lambda = \sin
\alpha' {\mathrm d} \alpha' {\mathrm d} \theta'$. We thus have
\begin{eqnarray}
E_s(\alpha)&=&\frac{f_0}{\alpha \sin \alpha }\int_\Omega
\frac{\alpha^2 \sin \alpha}{\alpha'^2 \sin \alpha' }
G\left(t_0,\phi,\lambda,f,\theta \right) \left| \cos \phi\right|
{\mathrm d}{\phi}{\mathrm d}{\lambda} \label{ES2c}\nonumber\\
\\
 &=&\frac{f_0}{\alpha \sin \alpha}\left\{ \int_{\mathrm{\Omega}}G\left(t_0,\phi,\lambda,f_0,\theta_0
\right)dA \right. \nonumber \\
& &  \times \left[1+O\left(\frac{\Delta_\alpha}{\alpha}\right)
\right]
\nonumber \\
 & &  +\int_\Omega  \left[G\left(t_0,\phi,\lambda,f,\theta
\right)-G\left(t_0,\phi,\lambda,f_0,\theta \right)\right]{\mathrm
d}A \nonumber
\\
& & \left. +\int_\Omega  \left[G\left(t_0,\phi,\lambda,f_0,\theta
\right)-G\left(t_0,\phi,\lambda,f_0,\theta_0
\right)\right]{\mathrm d}A \right\}  \label{ES2asym}. \nonumber\\
\end{eqnarray}
where  $\theta$ is the direction of the great circle at the
generation point that goes through that point and the observation
point. $\theta$ is thus a function of $\phi$, $\lambda$, $\alpha$
and $\theta_0$. $\Delta_\alpha$ is maximum value of $\left|\alpha'
- \alpha\right|$, i.e. the radius of the source region divided by
the Earth radius.

For continuous spectra, the second integral on the right hand side
of eq. (\ref{ES2}) goes to zero as $\alpha$ goes to infinity
(which, on the sphere is limited by $\pi$)  since the part of the
source spectra that contribute to the observations shrink to a
smaller and smaller neighborhood around $f_0$ and $\theta_0$. The
observed frequencies $f$ are limited by
\begin{equation}
\frac{\left|f-f_0\right|}{f_0} \leq \frac{\Delta_\alpha}{\alpha}
\end{equation}
This is enough to guarantee that this second integral also
contributes a deviation $\varepsilon_2$ from the asymptote,
limited by
\begin{equation}
\varepsilon_2 \leq A \frac{\Delta_\alpha}{\alpha} f_0\max
\left\{\frac{\partial G}{\partial f}\right\}\label{maxdF}
\end{equation}
where the maximum is taken over all the contributing components.

Similarly, the outgoing directions $\theta$ received at the
observation point are also limited to a narrow window as $\alpha$
increases, giving another deviation term $\varepsilon_3$. Using
the sine formula in the triangle $OPN$, $\sin \theta / \sin
(\pi-\alpha) = \sin \lambda/\sin \alpha'$. Thus, in the far field
of the storm and its antipode, $\theta$ is close to
$\theta_0=\lambda$. Thus $\sin \theta -\sin \theta_0=\sin \lambda
\left(\sin \alpha - \sin \alpha' \right)/ \sin \alpha'$, which is
less than $\Delta_\alpha/\sin \alpha'$, and therefore
$\left|\theta-\theta_0 \right|$ is less than $\arcsin
\left(\Delta_\alpha/ \sin \alpha'\right)$. If one does not get too
close to the storm or its antipode (say, $\Delta_\alpha < \alpha
<\pi - 2 \Delta_\alpha$) then  we can give an upper bound of
$1/\sin \alpha'$ as a function of $\alpha$ and obtain
\begin{equation}
\varepsilon_3  \leq A  \arcsin \left(2 \frac{\Delta_\alpha}{\sin
\alpha}\right) \max \left\{\frac{\partial G}{\partial
\theta}\right\}.
\end{equation}
The deviation from the asymptote due to $\varepsilon_3$ is thus of
the order of ${\Delta_\alpha}/{\sin \alpha}$ and may increase
close to the antipode, if waves from a wide range of directions
can reach that point. In practice this does not happen since
continents and island chains block most of the arrival directions
at the antipode, leaving only a small window of possible arrival
directions \citep[e.g.][]{Munk&al.1963}. Directional wave spectra
in active generation areas are generally relatively broad with
${\partial G}/{\partial \theta}/G$ typically less than 2 for
directions within 30$^\circ$ of the main wave direction. On the
contrary, typical storm spectra can give $f {\partial G}/{\partial
f}/G$ as large as 10 for frequencies within 30\% of the peak
frequency. We thus expect the deviation $\varepsilon_2$ to
dominate over $\varepsilon_3$.

\subsection{Verification}
\begin{figure}
\noindent\includegraphics[width=\columnwidth]{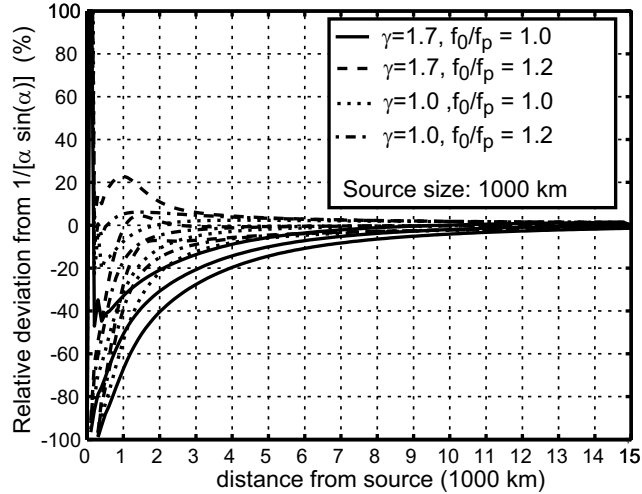}
\caption{Convergence of the swell energy $E_s$ integral
(\ref{ES2}) towards the asymptote $1/(\alpha \sin \alpha)$, as a
function of spectrum width for a storm diameter of 1000~km. The
result is independent of the choice of $f_p$. In practice the
calculations were made for $f_p=0.07$~Hz ($T_p=13$~s). The three
lines for each case correspond to position errors $\Delta_x$ of
-200, 0 and 200~km relative to the great circle trajectory. For
all cases considered here the deviation is less than 20\% beyond
4000~km from the storm center. \label{fig_asymptote}}
\end{figure}

For a  useful comparison with observations, the asymptotic swell
height evolution should be approached on a scale smaller than the
ocean basin scale. This is easily tested for storms with spatially
uniform spectra over a radius $r$,  by evaluating the integral in
eq. (\ref{ES2b}). We chose a center frequency $f_0$ and consider
the swell energy at a distance $R \alpha +\Delta_x$ and a time
$t(\alpha)$ such that $R \alpha =gt/(4 \pi f_0)$. $\Delta_x$ is
thus an error relative to the theoretical position of the wave
group of frequency $f_0$. The relative difference of $E_s(\alpha)$
and its asymptotic evolution $1/(\alpha \sin \alpha)$  depends
only on the spectral shape in the storm, the relative frequency
$f_0/f_p$ where $f_p$ is the peak frequency, the distance
$\alpha$, the storm size $\alpha / (r/R)$, and the position error
$\Delta_x$. Results are shown in figure \ref{fig_asymptote} for
isotropic directional spectra, in which case $\varepsilon_3=0$. Although an isotropic 
spectrum is not realistic at all, it allows for simple calculations. As discussed below, the effect of directional spreading is expected to be less important than the shape of the 
frequency spectrum. 

As indicated by eq. (\ref{maxdF}), the contribution of both the
spectral shape and $f_0/f_p$ comes through the maximum relative
variation of $F$ in the frequency interval that contribute to
$E_s$. Here we take the spectra in the storm to have a JONSWAP
shape \citep{JONSWAP}, that we adjust by varying the peak
enhancement factor $\gamma$. A relatively broad
\cite{Pierson&Moskowitz1964} spectrum is obtained with $\gamma=1$.
If we chose $f_0=f_p$, this spectrum will give smaller deviations
of $E_s$ from the asymptote (solid lines in figure
\ref{fig_asymptote}), than narrower spectra with larger values of
$\gamma$.  \cite{Young2006} showed that wave spectra in Hurricanes
generally fall in between these two categories, with a typical
value $\gamma_J=1.7$. Larger deviations from the asymptote are
obtained for $f_0 <  f_p$, since the forward face of the spectrum
is very steep, while smaller errors are obtained for $f_0
> f_p$ due to the more gentle decrease of $F$ towards the high
frequencies.

Similarly, large deviations are produced if the observations are
made in a direction far from the peak generation direction in
cases when the directional spectrum is narrow. For observation
directions 30$^\circ$ from the peak direction, and spectra with a
$cos^4$ directional distribution, the deviations are still
dominated by the dispersive term $\varepsilon_2$, as expected.

The other important factor is the distance $\alpha$ relative to
the storm size $\varphi_{\max}=r/R$. A faster convergence is
obtained for smaller storms. If observations do not correspond
exactly to a theoretical propagation at a group speed $C_g$ but
are within a distance $\Delta_x$ of the theoretical position, the
values of $E_s$ will also be affected in a way that depends on the
spectral shape. In the absence of energy gains or losses, and for
realistic storm sizes and spectral shapes, the deviation of
observations from the asymptote should be less than 20\% for
$x>4000$~km when $\Delta_x < 200$~km is enforced.

%
%
%

%
%
%
%

\begin{acknowledgments}
SAR data was provided by the European Space Agency (ESA) and buoy
data was kindly provided by the U.S. National Data Buoy Center,
and Marine Environmental Data Service of Canada. ESA funded the
XCOL co-located database and the initial work on the virtual buoy
concept. The swell decay analysis was funded by the French Navy as
part of the EPEL program. This work is a contribution to the
ANR-funded project HEXECO and DGA-funded project ECORS.
\end{acknowledgments}

%
%
%
%
%
%
%
%
%
%


\newcommand{\noopsort}[1]{} \newcommand{\printfirst}[2]{#1}
  \newcommand{\singleletter}[1]{#1} \newcommand{\switchargs}[2]{#2#1}


%
%

\end{article}




%
%
%
%
%
%


\end{document}